\documentstyle[pra,aps,epsfig,array,twocolumn]{revtex}
\begin{document}
\def\eps{\epsilon}
\newcommand{\ket}[1]{|  #1 \rangle}
\newcommand{\bra}[1]{ \langle #1  |}
\newcommand{\proj}[1]{\ket{#1}\bra{#1}}
\newcommand{\braket}[2]{\langle #1 | #2 \rangle}
\newcommand{\abs}[1]{ | \, #1 \,  |}
\newcommand{\av}[1]{\langle\,#1\,\rangle}
\newcommand{\asred}{\preceq}
\newcommand{\asredchem}{\leadsto}
\newcommand{\asequ}{\approx}
\newcommand{\exred}{\le}
\newcommand{\exredchem}{\to}
\newcommand{\exequ}{\equiv}
\newcommand{\exequchem}{\rightleftharpoons}
\newcommand{\stored}{\lesssim}
\newcommand{\tr}[2]{{\rm tr}_{\rm \scriptscriptstyle #1}(#2)}
\newcommand{\identity}{\mbox{\boldmath{1} \hspace{-0.33cm}
\boldmath{1} }}
\newcommand{\PT}[2]{(#1)^{\rm T_{#2}}}
\newcommand{\upp}[1]{^{\rm \scriptscriptstyle #1}}
\newcommand{\dnn}[1]{_{\rm \scriptscriptstyle #1}}
\newcommand\bea{\begin{eqnarray}}
\newcommand\eea{\end{eqnarray}}
\newcommand{\beq}{\begin{equation}}
\newcommand{\eeq}{\end{equation}}
\newtheorem{lem}{Lemma}
\newtheorem{theo}{Theorem}
\newtheorem{dfn}{Definition}
\newtheorem{cor}{Corollary}

\twocolumn[\hsize\textwidth\columnwidth\hsize\csname
@twocolumnfalse\endcsname
\title{Trust Enhancement by Multiple Random Beacons}
\author{Charles H. Bennett}
\author{John A. Smolin}
\address{IBM Research Division, Yorktown Heights, NY 10598,
USA --- {\tt bennetc, smolin@watson.ibm.com}}

\date{\today}
\maketitle
\begin{abstract}
Random beacons---information sources that broadcast a stream of
random digits unknown by anyone beforehand---are useful for
various cryptographic purposes.  But such beacons can be easily
and undetectably sabotaged, so that their output is known
beforehand by a dishonest party, who can use this information to
defeat the cryptographic protocols supposedly protected by the
beacon. We explore a strategy to reduce this hazard by combining
the outputs from several noninteracting (eg spacelike-separated)
beacons by XORing them together to produce a single digit stream
which is more trustworthy than any individual beacon, being random
and unpredictable if at least one of the contributing beacons is
honest. If the contributing beacons are not spacelike separated,
so that a dishonest beacon can overhear and adapt to earlier
outputs of other beacons, the beacons' trustworthiness can still
be enhanced to a lesser extent by a time sharing strategy.  We
point out some disadvantages of alternative trust amplification
methods based on one-way hash functions.
\end{abstract}
% close bracket for pretty quant-ph mode
]

\subsection{Introduction}
In cryptography and distributed computing, a random beacon is a
trusted information source (eg a radio transmitter) that
periodically broadcasts a random signal which is unknown to anyone
before the time of broadcast but becomes known to everyone
thereafter. Beacons were originally proposed by Rabin~\cite{R83}
to facilitate remote transactions such as contract signing.
Bennett, DiVincenzo and Linsker~\cite{BDL98} (cf.~Fig.~1) proposed
using a trusted random beacon to help authenticate video
recordings, made by untrusted recording apparatus operated by
untrusted personnel, against falsification of the time or content
(see Figure 1). These two applications require only a low
information rate (eg kHz), and assume that the history of
previously emitted signals becomes a matter of public record,
being stored at the beacon and/or other independent locations to
help resolve disputes. More recently, Aumann and Rabin~\cite{AR99}
have proposed using a much higher bandwidth beacon (eg GHz to THz)
to permit informationally secure encryption. The security of this
scheme depends on the beacon's information rate being so great
that no one can feasibly store the history of its previously
emitted signals.

The Achilles' heel of beacons is the need for users to trust that
they have not been sabotaged.  A dishonest beacon operator can
intentionally substitute pseudorandom digits, or true random
digits generated much earlier and leaked to accomplices, for the
supposedly fresh random digits being emitted by the beacon.  Even
if the operator is honest, a dishonest hardware supplier could
have concealed a tiny clandestine pseudorandom generator (PSRG) in
the supposed true random generator (TRG) hardware used by the
beacon, causing the hardware's output to be largely predictable.
To avoid detection, such a hardware saboteur should not make the
output wholly deterministic, because this would lead to the
sabotaged generator issuing the same digit stream \vbox{
\begin{figure}[htbf]
\epsfxsize=8.1cm \epsfbox{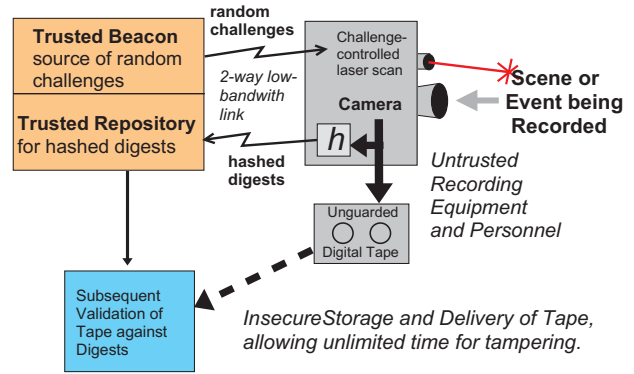} \medskip
\caption{Time-bracketed video authentication uses periodic
``challenge'' signals from a trusted random source to influence
the scene being recorded (e.g. by a challenge-controlled laser
scan), and shortly thereafter returns a hashed digest of the
scene, including the effect of the challenge, to a trusted
repository. The digests are produced by applying a secure hash
function $h$ to the current digital image data). The time
bracketing prevents pre- or post-dating, and provides evidence
that the action actually took place, as opposed to having been
computationally simulated in real time or assembled from
prerecorded material. Dishonest personnel can destroy the
videotape, or can prevent it from being recorded in the first
place, but so long as the beacon and repository remain honest,
they cannot easily produce a faked video that will match the
archived digests.}\label{Fig1}
\end{figure}
} \noindent every time it was turned on. Rather the sabotaged
generator might take its first few hundred digits from the TRG,
then use these as a seed for the concealed PSRG to generate the
rest of the sequence deterministically.  An accomplice, knowing
the nature of the sabotage, could then monitor the first few
hundred digits of beacon output and use these to predict all the
rest.  To help accomplices who had missed the initial beacon
output, the saboteur could use a steganographic reseeding
strategy, for example whenever a particular random 40-bit string
appeared in the beacon output, it would signal that the next 200
bits were not pseudorandom, but true random bits being used to
reseed the concealed PSRG.

 One might hope that these various sabotages, at least the ones
involving pseudorandom generators, could be detected by post-facto
analysis of the corrupted digit stream; but this hope is probably
vain, as it is widely believed that there exist
``cryptographically strong" pseudorandom generators which, when
seeded with a random $n$-bit seed, produce an output stream that
cannot be distinguished from true random digits in time polynomial
in $n$.

In view of the ease of sabotaging beacons and the difficulty of
detecting that they have been sabotaged, the main hope for beacon
users would appear to lie in protocols that amplify trust by
combining the outputs of several spatially and administratively
separated beacons, in the reasonable expectation that they only a
few of them have been sabotaged.  Henceforth we will consider a
set of $n$ nominally but not exactly synchronized beacons
$B_1...B_n$, each of which emits digits from an $\ell$ letter
alphabet at regular intervals.  Some beacons are are honest and
some dishonest (sabotaged), and we assume that the dishonest
subset does not change with time.  We will consider protocols for
trust amplification by users who have access to the outputs of all
the beacons.

An important consideration is whether the user, who combines the
output of several beacons to produce some resultant sequence, is
honest or dishonest. These two premises are profoundly different,
and give rise to quite different protocols. An honest user strives
to produce a resultant sequence that is random and unpredictable
by accomplices of the dishonest beacons, despite not knowing which
these are.  A dishonest user, by contrast, knows the identities of
the dishonest beacons, and conspires with them to produce a
predictable resultant sequence, despite the unpredictability of
the outputs of the honest beacons. A dishonest-user protocol is
considered successful if it defeats this conspiracy, forcing the
resultant sequence to be unpredictable even though the dishonest
user is trying to make it predictable. This is the relevant
premise for time-bracketed video authentication, whose goal is to
prevent a potentially dishonest camera manufacturer and operator
from producing a video that has been undetectably falsified as to
its time or content.

One might ask why an honest user needs any beacon at all: if he is
assumed to be honest, why can't he generate his own random
numbers, in effect being a beacon unto himself?  One possible
answer is that he may lack the physical capacity to produce random
numbers, or to produce them as fast as he desires, without drawing
on external sources of randomness. In passing we note that an
honest user, having a low-rate random source in his own lab, can
use an {\em extractor\/}~(cf.~\cite{NT99}, to distill certifiably
unpredictable high-rate random numbers from the low-rate private
source and a collection of high-rate random beacons, only some of
which are honest.

The remainder of this paper will concern dishonest-user protocols
of the kind relevant for beacons to be used in time-bracketed
authentication.

Vazirani~\cite{V87} considered the related problem of devising
protocols to extract nearly unbiased random bits from two beacons,
{\em both\/} dishonest and colluding, but neither entirely
controllable by its operator.   Here, by contrast, we have some
beacons that are entirely controllable by colluding dishonest
operators, and others that are entirely random and honest, but the
designer of the protocol doesn't know which.

\subsection{Trust amplification for beacons that are spacelike separated
or otherwise known to be incapable of influencing one another}
Trust amplification works best when the beacons are known to be
incapable of influencing one another, so the dishonest beacons
cannot adapt their output to that of the honest ones. This will be
assured if the beacons' emissions are so well synchronized,
compared to the distance between them, as to be spacelike
separated in the sense of special relativity.  Two beacons, say
$B_1$ and $B_2$, are said to be spacelike separated when for all
integer $i$ the spacetime event ${\cal E}(i,B_1)$ occurs at a
spacelike interval from the spacetime event ${\cal E}(i,B_2)$; in
other words, a light signal starting at beacon $B_1$ at the
instant when it emits its $i'$th digit $B_1(i)$ will not have
arrived at beacon $B_2$ by the time beacon $B_2$ emits its $i$'th
digit $B_2(i)$, and similarly with the indices 2 and 1 reversed.
Under these conditions, it is evident that the  XOR (or more
generally the mod-$\ell$ sum $\oplus$, for an $\ell$-letter
alphabet) of the two beacons, ie the digit stream $B_1(i)\oplus
B_2(i)$, will be random iff at least one of constituent beacons is
random.~\footnote{Even without spacelike separation,
noninteraction can sometimes be assured, with lesser confidence,
by shielding or isolating each potentially dishonest beacon well
enough to block significant incoming signals from the honest
beacons.  However, unless otherwise noted, we will henceforth
assume that the beacons are not shielded or isolated, and that
unless two beacons are spacelike separated the later one, if
dishonest, can overhear and adapt to the signals of the earlier
one.}

For a beacon to be trustworthy, its output must not only be
random, but also unpredictable before the time it is supposed to
become public.  In general the information from any given beacon
does not become available to everyone simultaneously, owing to
propagation delays, which can never be less than the distance of
the observer from the beacon divided by the velocity of light $c$.
For example, in the case of two synchronized honest beacons
separated by distance $d$, an observer midway between the beacons
would learn $B_1(i)$ and $B_2(i)$, and could compute $R_{XOR}(i)$,
at a time $d/2v$ after the emission time of the $i$'th digit,
where $v\leq c$ is the signal propagation velocity. An observer at
either beacon would have to wait a little longer, until time
$d/v$, to obtain the signal from the other beacon. These
considerations may be summarized in the following proposition,
whose proof is obvious:

\vbox{ If $B_1...B_n$ is a set of spacelike separated beacons, at
least one of which is honest,

\begin{enumerate}
\item the modular sum
\beq R_{XOR}(i)=\bigoplus_{k=1}^n B_k(i) \eeq is random,

\item $R_{XOR}(i)$ is unpredictable from the viewpoint of any observer
outside the intersection of the forward light cones of the honest
subset of beacons.

\item assuming that signals propagate at light speed, $R_{XOR}(i)$ can be
correctly calculated by any observer inside the intersection of
the forward light cones of all the beacons. \end{enumerate} } The
principal effect of dishonesty is thus to create a region of
spacetime within which $R_{XOR}(i)$ is predictable to accomplices
of the dishonest beacons, but not to the general public.  This
region consists of points within the forward light cone of every
honest beacon, but outside the forward light cone of at least one
dishonest beacon.

\subsection{Trust amplification for beacons that are timelike separated
or otherwise suspected of influencing one another} Within any
nominally synchronized set of beacons there may be enough timing
error that the beacons are not in fact spacelike separated. Lack
of spacelike separation can seriously impair the trustworthiness
of the resultant sequence $R_{XOR}(i)$, making it untrustworthy
over all spacetime, not just in a limited region. For example,
suppose that beacon $B_1$ is so late that it has all the other
beacons in its past light cone. Then, if $B_1$ is sabotaged, it
can adapt its output $B_1(i)$ so as to force the resultant
$R_{XOR}(i)$ not to be random, but to take on a predetermined
value, perhaps chosen long beforehand. Thus the accomplices of the
dishonest beacon potentially know $R_{XOR}(i)$ wherever they sit
in spacetime, while honest players, as before, will only know
$R_{XOR}(i)$ if they sit within the intersection of the future
light cones of all the beacons, which in this case is simply the
future light cone of $B_1$.

In the worst case, where one beacon is consistently so late as to
have all the others in its past light cone, the XOR of all the
beacons is no more trustworthy than the single latest beacon taken
by itself. However, one can still gain some increased trust by
combining the beacons in a different fashion, which we call the
time-sharing protocol.  Here the resultant is defined to be a
cyclicly chosen one of the original beacons,

\beq R_{TS}(i)= B_{i \,{\rm mod}\, n}(i). \eeq

If some of the beacons are honest and some dishonest, then some
digits of the resultant sequence $R_{TS}$ will be predictable by
accomplices of the dishonest beacons and others will be
unpredictable. The resultant sequence is thus sure to be partly
unpredictable, while the sequence from any individual beacon, or
the XOR of all of them, has some chance of being wholly
predictable. For purposes such as time-bracketed
authentication~\cite{BDL98}, a sequence that is sure to be at
least partly unpredictable is still usable, though not as good as
a wholly unpredictable sequence; but a sequence that has some
chance of being be wholly predictable is unusable.

The sort of uncertain unpredictability relevant to time-bracketed
authentication can be quantified by the per-character {\em min
entropy}, ie the logarithm of the probability (as seen by the
dishonest users) of the most likely resultant sequence
$R_{TS}(i)$, divided by the length of the sequence. If there are
$n$ timelike separated beacons, $k$ of which at random are
sabotaged but we don't know which, then the min entropy of
$R_{TS}$ is $((n-k)/n)\log\ell$ bits per character. On the other
hand, each individual beacon, say $B_1$, or the XOR of all the
beacons if $B_1$ is the latest, has a per-character min entropy
approaching zero, because $B_1$'s min entropy is dominated by the
probability $k/n$ that it is sabotaged, and so emits a sequence
that is completely predictable by dishonest users.

The advantage of using min entropy can be seen by noting that in
this situation the ordinary Shannon entropies of $R_{TS}$ and
$B_1$ are equal, both being $((n-k)/n)\log\ell$  bits per
character. Thus min entropy heavily and properly penalizes any
chance of complete predictability, while Shannon entropy allows it
to hide amidst the unpredictability of other cases.

The superiority of spacelike separation, and the advantage of
using the $R_{XOR}$ instead of $R_{TS}$ when the beacons are known
to be spacelike separated, can be seen by comparing the per
character min entropies in various cases.

\bigskip
\vbox{ {\small
\begin{table}
\begin{center}
\begin{tabular}{c|c|c}

beacon separation & spacelike &  timelike \\
\hline
XOR protocol &  1& 0\\
\hline
Time sharing protocol  &  $(n\!-\!k)/n$      &  $(n\!-\!k)/n$ \\

\end{tabular}\end{center} Table I. Per character min
entropy of resultant sequences $R_{XOR}$ and $R_{TS}$ obtained
respectively by XOR and time-sharing protocols for trust
amplification. We assume $n$ beacons, an unknown $k$ of which are
dishonest. Entropies are in units of $\log\ell$, the entropy of an
honest beacon emitting characters from an $\ell$-letter alphabet.
\end{table} }
}

In general the resultant sequences $R_{XOR}$ (for spacelike
separated beacons) or $R_{TS}$ (for any set of beacons) will be at
least partly unpredictable, and therefore usable for purposes such
as time bracketed authentication, except when all the beacons are
dishonest.

\subsection{Why it is not generally advantageous to combine
beacons by hashing} A seemingly attractive alternative to
$R_{XOR}$ and $R_{TS}$ would be to use a cryptographically strong,
one-way hash function $h$ to combine the beacons,eg
 \beq R_h(i)=h(B_1(i),B_2(i)...B_n(i)), \eeq
where $h$ is an $m$-to-$1$ mapping on characters from an
$\ell$-letter alphabet; but, as we shall show, $R_h$ has
significant weaknesses compared to $R_{XOR}$ and $R_{TS}$.  For
concreteness consider the case where there are $m=2$ beacons, an
unknown one of which is dishonest, and where each beacon
broadcasts letters from an $\ell=2^d$ letter alphabet, so that
$R_h$ may be viewed as a pseudorandom mapping from a pairs of
$d$-bit strings to a single $d$-bit string.  In the following we
will use lower case letters $x,y$ etc to denote $d$-bit strings.

If the two beacons are spacelike separated, or otherwise known to
be noninteracting, $R_{XOR}$ will perfectly random and
unpredictable.  By contrast, as we will show, a dishonest beacon
operator with a lot of computing power (whom we will call Eve) can
force $R_h$ to be significantly nonrandom, for example forcing its
first bit to be almost always zero. Assume, without loss of
generality, that Eve is operating $B_1$.  She then finds, by brute
force search, some string $x^\dagger$ defined to be any $d$-bit
string $x$ on which the set $\{y:h(x,y) \;{\tt begins}\;{\tt
with}\; 1 \}$ has minimal cardinality. This minimal cardinality
will be of order unity, so if Eve always broadcasts $x^\dagger$ as
her maliciously chosen output from $B_1$, while the honest beacon
$B_2$ broadcasts random $d$-bit strings, the first bit of
$R_h(x^\dagger,y)$ will almost always be 0. Similarly Eve can
force the value of any other single digit of $R_h$, or strongly
bias several digits of her choosing.  Under the more realistic
assumption that Eve has limited computing power, she cannot
appreciably bias $R_h$, so it is no better and no worse than
$R_{XOR}$.

If the beacons are timelike separated, with dishonest beacon $B_1$
later so it can overhear honest beacon $B_2$, a computationally
powerful Eve can force the {\em almost all\/} the digits of $h$ to
agree with a particular string $z_0$ of her choosing.  To do so,
she waits till she has heard the particular string $y$ broadcast
by $B_2$, then chooses her string $x^\ddag$ to be one that
minimizes the Hamming distance between $z_0$ and $h(x,y)$.  Often
(about $1/e$ of the time) she can obtain an exact hit
$h(x^\ddag,y)=z_0$; in other case she can almost certainly find an
$x^\ddag$ for which $h(x^\ddag,y)$ differs from the target $z_0$
in only one bit.  Under the more realistic assumption of limited
computing power, Eve can force the values of $m\approx$ several
dozen bits of her choosing in $R_h$.  To do so she waits till she
overhears $y$, then proceeds by trial and error, evaluating
$h(x,y)$ for $2^m$ random $x$ values, until she finds one, $x^*$
such that the first $m$ digits, or any other set of $m$ digits of
her choosing in $R_h(x^*,y)$, have the values she wants.  The
computational effort is exponential in the number of digits she
wishes to force.   For small string lengths $d$, an Eve with
moderate computing power can force all the digits of $R_h$, making
it totally insecure.  For large $d$, this is no longer possible,
and $R_{TS}$ and $R_h$ offer two somewhat different kinds of
partial unpredictability.  With $R_h$ such an Eve can force a
small fraction of the digits in the output stream, digits of her
choosing, while the others remain unpredictable.  With with
$R_{TS}$ Eve can force half the digits in the output stream, but
has no control over which digits these are. Depending on the
beacon application, one or another kind of partial predictability
may be preferable.  For time-bracketed authentication, it is
probably better to use $R_{TS}$, because a few absolutely
uncontrollable challenges are probably harder for a would-be
forger to simulate than a greater number of partly-controllable
challenges.


\begin{references}


\bibitem{R83}M.O. Rabin ``Transaction Protection by Beacons",
J.Computer and System Sciences 27(2): 256-267 (1983)

\bibitem{BDL98} C.H. Bennett, D.P. DiVincenzo, and R. Linsker,
``Digital recording system with time-bracketed authentication by
on-line challenges and method for authenticating recordings" US
Patent 5764769  (1998 Further details at
www.research.ibm.com/people/b/bennetc/AVC.ppt ).

\bibitem{AR99} Y.Aumann, and M.O. Rabin ``Information theoretically
secure communication in the limited storage space model",
CRYPTO1999: 65-79 (1999).

\bibitem{V87} U. Vazirani ``Towards a Strong Communication Complexity Theory
or Generating Quasi-Random Sequences from Two Communicating
 Semi-Random Sources'' Combinatorica, {\bf 7(4),} 375-392 (1987)

\bibitem{NT99} N. Nisan, A. Ta-Shma, ``Extracting Randomness: A
Survey and New Constructions'' J.C.S.S. {\bf 58}, 148-173 (1999)

\end{references}
\end{document}